\documentclass[aps,twocolumn,pra,tightenlines,floatfix,showpacs,superscriptaddress,longbibliography]{revtex4-1}
\usepackage{graphicx}
\usepackage{amsmath}
\usepackage{amssymb}
\usepackage{times}
\usepackage{epstopdf}
\usepackage[english]{babel}
\usepackage{natbib}
\usepackage{hyperref}
\hypersetup{colorlinks=true, citecolor=blue, linkcolor=blue}
\usepackage{color}
\usepackage{cleveref}
\usepackage{tablefootnote}

\crefformat{equation}{Eq.~(#2#1#3)}
\crefmultiformat{equation}{Eqs.~(#2#1#3)}{ and~(#2#1#3)}{, (#2#1#3)}{ and~(#2#1#3)}
\Crefformat{equation}{Equation~(#2#1#3)}
\crefformat{section}{Sec.~#2#1#3}
\Crefformat{section}{Section~#2#1#3}
\crefformat{figure}{Fig.~#2#1#3}
\Crefformat{figure}{Figure~#2#1#3}

\newcommand{\bra}[1]{\ensuremath{\left\langle#1\right|}}
\newcommand{\ket}[1]{\ensuremath{\left|#1\right\rangle}}
\newcommand{\tr}{\ensuremath{\textrm{tr}}}
\newcommand{\dg}{\ensuremath{^\dagger}}

\newcommand{\ie}{\emph{i.e.,}~}
\newcommand{\eg}{\emph{e.g.,}~}

\begin{document}
\title{Engineered thermalization and cooling of quantum many-body systems}
\author{Mekena Metcalf}
\affiliation{Computational Research Division, Lawrence Berkeley National Laboratory, Berkeley, CA 94720, USA}
\email{mmetcalf@lbl.gov}
\author{Jonathan E. Moussa}
\affiliation{Molecular Sciences Software Institute, Blacksburg, VA 24060, USA}
\author{Wibe A. de Jong}
\affiliation{Computational Research Division, Lawrence Berkeley National Laboratory, Berkeley, CA 94720, USA}
\author{Mohan Sarovar}
\affiliation{Extreme-scale Data Science and Analytics, Sandia National Laboratories, Livermore, CA 94550, USA}
\email{mnsarov@sandia.gov}

\begin{abstract}
We develop a scheme for engineering genuine thermal states in analog quantum simulation platforms by coupling local degrees of freedom to driven, dissipative ancilla pseudospins. We demonstrate the scheme in a many-body quantum spin lattice simulation setting.  A Born-Markov master equation describing the dynamics of the many-body system is developed, and we show that if the ancilla energies are periodically modulated, with a carefully chosen hierarchy of timescales, one can effectively thermalize the many-body system. Through analysis of the time-dependent dynamical generator, we determine the conditions under which the true thermal state is an approximate dynamical fixed point for general system Hamiltonians. Finally, we evaluate the thermalization protocol through numerical simulation and discuss prospects for implementation on current quantum simulation hardware. 
\end{abstract}

\maketitle

\section{Introduction}

Preparation of mixed states of many-body systems, particularly thermal states at low temperatures, is valuable for many scientific and algorithmic tasks, \eg calculating finite temperature response of materials, Gibbs sampling for machine learning and optimization \cite{brandao_quantum_2016, Kliesch_2011, Riera_2012}. 
However, computing, or sampling from, such low temperature thermal states of large quantum (or classical) many-body systems is a notoriously difficult problem with a long history \cite{binder_monte_2010,landau_guide_2014,pang_introduction_2016}.
Analog quantum simulators, controllable experimental platforms that can be engineered to mimic and simulate quantum many-body systems \cite{QuantumSimRev}, present a new approach for preparing and sampling from complex states of such systems, including low-temperature thermal states. 
Despite this, most analog quantum simulation platforms typically focus on preparing pure quantum states of many-body systems  . This is partly because preparing mixed states on many leading analog quantum simulation platforms, \eg trapped cold atoms, is challenging due to the lack of scattering mechanisms needed to dissipate energy and thermalize \cite{Bloch_UCARev_2017, MetcalfPC}. Therefore, the thermalizing environment must be also engineered and simulated, which is challenging since such environments typically contain an immense number of degrees of freedom.

We develop a technique to engineer thermalization of many-body spin Hamiltonians based on coupling to driven, dissipative ancilla degrees of freedom (DOF) that effectively act as a {\it tunable macroscopic bath}. 
We derive a Born-Markov master equation describing the dynamics of a many-body system coupled to fast-relaxing, driven ancilla qubits, and show that if the ancilla energies are periodically modulated and swept across the system energy spectrum, with a carefully chosen hierarchy of timescales, one can effectively thermalize a many-body system. Combing the spectrum with ancillary spins has been proposed for ground-state cooling on digital quantum simulation platforms \cite{kaplan_ground_2017, DigitalCooling}, we extend these ideas by coupling ancilla driven to thermal equilibrium. We use analytic arguments and numerical investigations to demonstrate that the true thermal state is an approximate fixed point of the dynamics. The scheme we develop can be viewed as a protocol for filtering and transforming the structureless electromagnetic vacuum reservoir to a structured reservoir suitable for thermalizing the many-body quantum system at hand. This enables thermalization using a finite number of controlled ancilla DOF.

Our development of this thermalization protocol builds upon previous work that examined techniques for thermalizing systems governed by stabilizer Hamiltonians \cite{herdman_stroboscopic_2010, young_finite_2012}. In these works, it was shown that many-body systems governed by a stabilizer Hamiltonians can be driven to their thermal states by weakly coupling driven and dissipated ancilla DOF. When one attempts to generalize the constructions in these works to achieve the goal of thermalizing an arbitrary many-body system, several issues arise. Stabilizer Hamiltonians possess local excitations, and therefore local couplings to ancilla DOF suffice to provide energy excitation and dampening. These Hamiltonians are harmonic in the sense that eigenstates within an excitation sector are separated by a constant (and known) energy -- \eg for the toric code, each quasiparticle excitation adds a known constant energy, depending only on whether it is a electric charge or magnetic vortex   \cite{kitaev_topological_2009}. This implies that the ancilla DOFs that induce excitations can be tuned to a single energy (or few energies) to induce energy-conserving transitions between system and ancilla. Consequently, if Boltzmann populations are maintained in the ancilla DOFs, one can guarantee transition dynamics within the system that obey detailed balance. These observations are at the core of the stabilizer Hamiltonian thermalization protocol constructed in \cite{young_finite_2012}.

Systems governed by general Hamiltonians possess non-local excitations and non-uniform spectra. Therefore, one does not expect the stabilizer thermalization protocol to translate to a more general setting. However, we will show that by introducing a \emph{time-dependent}, driven set of ancilla DOF one can formulate a slightly modified protocol that thermalizes systems governed by non-stabilizer Hamiltonians. 

Our work is also motivated by the recent work by Shabani and Neven \cite{shabani_artificial_2016} that studied how to engineer a reservoir to achieve thermalization of a quantum many-body system. gTheir constructions were based on approximating Kubo-Martin-Schwinger (KMS) conditions for equilibration, which the authors show can be achieved by suitably driving a large number of independent harmonic oscillators that  together form an engineered reservoir. In this work we set out to demonstrate that thermalization is also possible with a reservoir composed of a small number of ancilla qudits (finite dimensional systems) when coupled with time-dependent driving and local dissipation. This significantly increases the practicality of the engineered thermalization scheme. It is important to point out that we do not make any guarantees on the thermalization time of our protocol. We expect that this will diverge with system size and inverse temperature for systems governed by Hamiltonians with hard to prepare ground states, such as QMA-hard Hamiltonians \cite{kempe_complexity_2006, aharonov_power_2009}.

Finally, we note that there exist alternative thermalization protocols that exploit the Eigenstate Thermalization Hypothesis (ETH) and sample a subsystem of a larger system \cite{Zapata_Boltzmann_2019}. The state of this subsystem could be described by a thermal state if the ETH holds. These protocols do not possess a fully-controllable macroscopic bath and temperature must be determined by post-selection, therefore, it would be unlikely these methods can generate specific thermal states efficiently. In contrast our protocol does not require post-selection since the state of the many-body system converges to the thermal state. 

The remainder of this paper is structured as follows.  \Cref{sec:model} presents the general model, thermalization protocol and discusses the relevant parameters. 
\Cref{sec:reductions} derives a reduced equation of motion by averaging over the dynamics of the ancilla DOF in a carefully considered parameter regime.  Then in \cref{sec:db} we discuss the importance of the detailed balance condition in dictating the accuracy of our thermalization protocol, and present analytic arguments for the properties of the steady state of our engineered evolution. \cref{sec:sims} presents several numerical simulation results that serve to illustrate our protocol and highlight some of its key properties, including its lack of accuracy in some regimes. Then we conclude in \cref{sec:concl} with a brief discussion.

\begin{figure}[t!]
	\includegraphics[width = 3.4in] {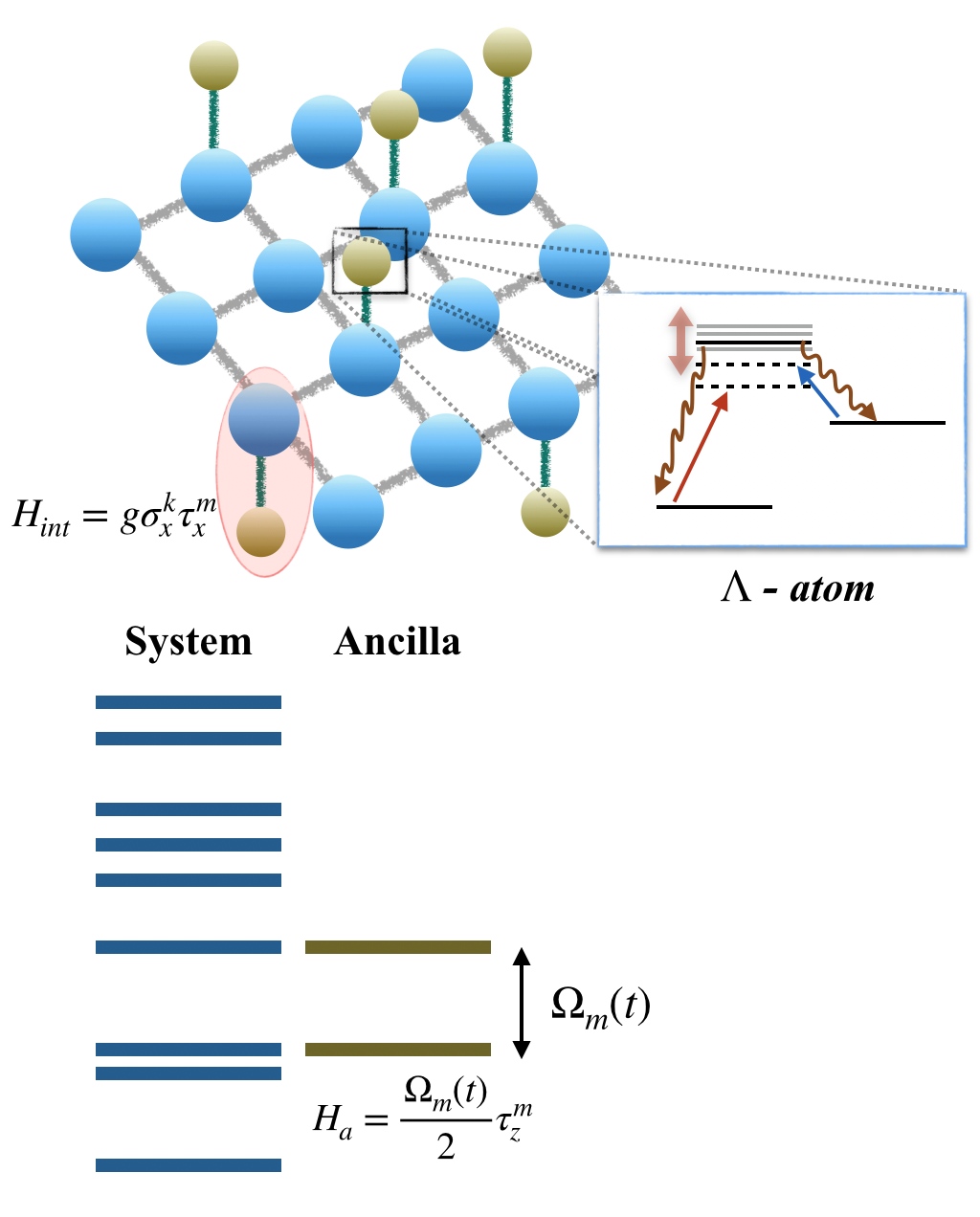}
	\caption{Principal spins (blue) composing a lattice system are coupled to optically pumped, $\Lambda$-atoms (yellow). Two time-dependent levels of the $\Lambda$-atoms are coupled to the principal spins by an exchange interaction. The populations of these two levels are maintained at Boltzmann distributions through optical pumping, and their energies are swept across the principal system's spectrum.}
	\label{Fig:Schematic}
\end{figure}

\section{The model}
\label{sec:model}
Consider a many-body Hamiltonian, $H_{\rm sys}$ describing coupled quantum spins. We will refer to the spin DOF governed by this Hamiltonian as the \emph{principal} spins, and these are often localized to a regular lattice, see Fig. \ref{Fig:Schematic}. We denote the eigenvalues and eigenvectors of $H_{\rm sys}$ by $\epsilon$ and $\ket{\epsilon}$, respectively.

Some subset of these principal spins are coupled to independent ancillary DOF that can also be described as quantum two-level systems, which we will refer to as ancilla spins. The coherent dynamics of the state of the combined system, living in the Hilbert space $\mathcal{H}_{sys}\otimes \mathcal{H}_{ancilla}$, is given by the time-dependent Hamiltonian ($\hbar=1$)
\begin{equation}
    H(t) = H_{\rm sys} - \sum_{m=1}^M \frac{\Omega_m(t)}{2}\tau_z^m + \sum_{m=1}^M g_m \sigma_\alpha^{k_m}\tau_x^m, 
    \label{eq:full_ham}
\end{equation}
where $\tau_\alpha^m$ ($\alpha = x,y,z$) define Pauli spin operators on ancilla spin $m$, and $\sigma_\alpha^k$ define Pauli operations on the $k^{\rm th}$ principal spin. The ancilla spin energies are time-dependent, and we assume this dependence is periodic and takes the form $\Omega_m(t) = \Delta f_m(t)$ where $\Delta$ is the difference between the system Hamiltonian's largest and smallest energy, and $f_m(t)$ is a periodic function. The ancilla are coupled weakly to the system by an excitation exchange interaction between system spin $k_m$ and ancilla $m$ with coupling strength $g_m$. 
Weak coupling of the system and bath is required for thermalization in the original eigenbasis of $H_{\rm sys}$. In addition to this Hamiltonian, the ancilla spins must be driven to thermal equilibrium on a faster time-scale than the system-ancilla coupling timescale. We assume that each ancilla spin is driven (damped) at a rate $\gamma^m_+$ ($\gamma_-^m$) such that it effectively evolves according to the standard Markovian master equation:
\begin{align}
     \frac{d\rho^m}{dt} &= \mathcal{L}_m(t)\rho^m(t)\\
     &= \gamma_+^m(t)\mathcal{D}[\tau_+^m]\rho^m(t) +\gamma_-^m(t)\mathcal{D}[\tau_-^m]\rho^m(t),
     \label{eq:optical_pumping_me}
\end{align}
where $\rho^m(t)$ is the density matrix for ancilla spin $m$  and $\mathcal{D}[A](\cdot) = A(\cdot)A\dg - 1/2\{A^\dagger A,(\cdot)\}$ with the Pauli raising (lowering) operator $\tau_+^m$($\tau_-^m$). In the following, we will choose 
\begin{equation}\label{Eqn:BathDB}
    \frac{\gamma_-^m(t)}{\gamma_+^m(t)} = e^{\beta\Omega_m(t)},
\end{equation}
 with $\beta = \frac{1}{T_{\rm eff}}$ being the inverse temperature that we wish to thermalize the principal spin system to. Further, we choose the pumping parameter $\Gamma^m \equiv \gamma_+^m + \gamma_-^m \gg \vert \frac{d\Omega_m(t)}{dt} \vert, \forall t$. In this regime the quasi-static fixed point of the ancilla spin's evolution (which it reaches in roughly $1/\Gamma^m$ time) is a density matrix with Boltzmann distributed populations with respect to the inverse temperature and energy $\Omega_m(t)$. One can engineer such an open-system evolution of a two-level system by encoding the two levels within a driven and dissipated (\emph{e.g.,} optically pumped) $\Lambda$-atom \cite[Chap. 7.9]{steck_quantum_2013}, see Fig.~\ref{Fig:Schematic}. Importantly, direct dissipation of the principal spins does not lead to generation of a thermal state of the many-body system since the steady state in this case is dominated by locally thermalized spins (in a product state).

The number of ancilla ($M$) required and the choice of system-ancilla coupling (\ie the value of $\alpha$ and $k_m$ in each coupling $\sigma_{\alpha}^{k_m}$) is set by the requirement of ergodicity of dynamics of the system. Intuitively, this requirement states that it should be possible to transition from any eigenstate of $H_{\rm sys}$ to any other eigenstate through application of a sequence of ancilla-induced operations. We will state a concrete algebraic condition for ergodicity in the following section that can easily be checked. In practice, we have found that ergodicity is satisfied with at most $M=N$ ancilla spins (as long as the system-ancilla interactions do not commute with the system Hamiltonian).

Our thermalization protocol, which aims to drive the collective state of the principal spins to the Gibbs state under $H_{\rm sys}$ at inverse temperature $\beta$, \emph{i.e.,} $\propto e^{-\beta H_{\rm sys}}$,  proceeds by engineering the above interactions between the principal spins and ancilla spins and then sweeping all 
$\Omega_m(t)$ over a small number of periods. The intuition for why this should thermalize the principal system comes from noticing that at specified times, some set of ancilla spins is resonant with some energy gaps in $H_{\rm sys}$, and since the ancilla spin populations are Boltzmann distributed, the interaction will drive populations in the resonant energy levels towards being Boltzmann distributed. Note that in this work we use the term ``gap'' to refer to any transition energy in the system, not just an energy difference between neighboring states. Over time and several cycles of the periodic modulation, as the ancilla spins become resonant with more and more energy gaps in $H_{\rm sys}$ all system populations will be driven to their thermal equilibrium populations. This intuitive description is what we formalize in the following sections. 

\section{Reduced description of dynamics}
\label{sec:reductions}

Understanding the steady-state properties of the time-dependent evolution prescribed in the previous section is not possible without some simplifications. In order to gain insight into the evolution of the system under the protocol we find a particular parameter regime in which we can average over the ancilla spins and derive a time-dependent Markovian master equation governing the evolution of the principal spins alone. 

In order to thermalize the system we need to be in a regime where the system ``sees'' each ancilla energy for some time in order to exchange energy (at the engineered rates). In addition, we do not want to couple the ancilla spins too strongly to the system, otherwise one cannot guarantee thermalization in the original eigenbasis of $H_{\rm sys}$. These considerations lead us to require the parameter regime choice:
\begin{align}
	\vert\frac{df_m(t)}{dt}\vert \ll g_m \lessapprox \Gamma^m  \ll ||H_{\rm sys}||, ~~~ \forall m,t
	\label{eq:regime}
\end{align}

As we prove in the Appendix, within this parameter regime, we can derive a time-dependent, Markovian master equation describing the dynamics of the principal spins alone (in an interaction picture with respect to the free Hamiltonian of the spin lattice, $H_{\rm sys}$):
\begin{widetext}
\begin{equation}
    \frac{d\rho(t)}{dt} = \sum_{m=1}^M g_m^2\sum_\omega \lambda^m_t(\omega)\left[ X_m(\omega)\rho(t)X_m^\dagger(\omega) - \frac{1}{2}X_m^\dagger(\omega)X_m(\omega)\rho(t) - \frac{1}{2}\rho(t)X_m^\dagger(\omega)X_m(\omega)\right].
    \label{eq:L}
\end{equation}
\end{widetext}
Here, $X(\omega)$ are {\it time-independent}, frequency resolved ancilla coupling operators on the system
\begin{align}
	X_m(\omega) = \sum_{\{\epsilon',\epsilon | \epsilon'-\epsilon = \omega\}}\Pi(\epsilon) \sigma_\alpha^{k_m} \Pi(\epsilon'),
\end{align}
where $\Pi(\epsilon)$ is a projector on the eigenspace of $H_{\rm sys}$ with eigenvalue $\epsilon$. $\omega>0$ denotes downward transitions that decrease the energy of the system, and $\omega<0$ denotes upward transitions that increase system energy. The coefficients $\lambda^m_t(\omega)$ are the time-dependent spectral densities the principal spins experience as a result of the ancilla dynamics (derived from the ancilla correlation functions in Appendix~\ref{app:derivation}), and can be thought of as specifying the rates of downward and upward (in system energy) transitions. They take Lorentzian form:
\begin{align}
\lambda^m_t(\omega) = \frac{P^m_t \left(\frac{\Gamma^m}{2}\right)}{\left(\frac{\Gamma^m}{2}\right)^2 + (\omega - \Omega_m(t))^2} + \frac{(1-P^m_t) \left(\frac{\Gamma^m}{2}\right)}{\left(\frac{\Gamma^m}{2}\right)^2 + (\omega + \Omega_m(t))^2},
\label{eq:lambda}
\end{align}
with $P^m_t = e^{\beta \frac{\Omega_m(t)}{2}}/(e^{\beta \frac{\Omega_m(t)}{2}}+e^{-\beta \frac{\Omega_m(t)}{2}})$ being the ground state Gibbs population of the ancilla with energy splitting defined by $\Omega_m(t)$. 
We have assumed here for simplicity that the dampening parameter, $\Gamma^m$, is time-independent for all ancilla.
Note that due to the Lorentzian form, these rates are significant only when $\pm\omega \approx \Omega_m(t)$; \ie around frequencies in resonance with the ancilla energies. We also note that $\lambda_t^m(\omega) \geq \lambda_t^m(-\omega)$ for $\omega>0$, confirming that at any finite temperature the downward transition rates dominate over upward rates.

As mentioned above, in order to thermalize using such ancilla-driven dynamics we must ensure that the system-ancilla couplings generate ergodic system dynamics. Ergodicity of system dynamics can be checked by evaluating the well-known algebraic sufficient condition stated in terms of the frequency resolved Lindblad operators \cite{breuer_theory_2002}:
\begin{align}
[K, X_m(\omega)]=0 \quad \forall ~m, \omega \implies K \propto I.
\end{align}
In other words, the commutant of the set of frequency resolved Lindblad operators $X_m(\omega)$ is trivial. In the following, we will assume that the number and types of system-ancilla couplings have been chosen to satisfy this condition.  

\section{Detailed balance}
\label{sec:db}
What are sufficient conditions for the thermal state of the principal spins, 
\begin{align}
	\rho_\beta = \frac{e^{-\beta H_{\rm sys}}}{\tr(e^{-\beta H_{\rm sys}})}
\end{align}
to be the steady state of this evolution? To answer this question, we evaluate the fixed point of the generator of evolution in \cref{eq:L}. This follows the standard analysis of fixed points of Lindblad generators, \eg \cite[Ch. 3.3]{breuer_theory_2002}. 

We start by listing some identities that are easily derived from the definitions of the relevant operators:
\begin{subequations}
    \begin{equation}
        \left[H_{\rm sys},X_m(\omega)\right] = -\omega X_m(\omega)
    \end{equation}
    \begin{equation}
        \left[H_{\rm sys},X_m^\dagger(\omega)\right] = \omega X_m^\dagger(\omega)
    \end{equation}
    \begin{equation}
        \rho_\beta X_m(\omega) = e^{\beta\omega}X_m(\omega)\rho_\beta
    \end{equation}
    \begin{equation}
        \rho_\beta X_m^\dagger(\omega) = e^{-\beta\omega}X_m^\dagger(\omega)\rho_\beta.
    \end{equation}
\end{subequations}
Using these properties, and the fact that $X_m^\dagger(\omega) = X_m(-\omega)$ \footnote{This property relies on the system operator of the system-bath coupling ($\sigma_\alpha$ in \cref{eq:full_ham}) being Hermitian. However, this derivation can be generalized to cases where the coupling is not Hermitian also.}, we write $d\rho_\beta/dt$ as:
\begin{align}
\frac{d\rho_\beta}{dt} 
	&= \sum_{m=1}^M g_m^2 \sum_\omega \left[ \lambda^m_t(\omega) - \lambda^m_t(-\omega)e^{\beta \omega}\right] \times \nonumber \\
	& \quad\quad X_m(\omega)\rho_\beta X_m^\dagger(\omega),
	\label{eq:tdep_gibbs}
\end{align}

This expression yields a sufficient condition for the system thermal state being a fixed point of the engineered evolution; \ie $d\rho_\beta/dt=0$ if
\begin{align}
	\lambda^m_t(\omega) - \lambda^m_t(-\omega)e^{\beta \omega}=0, \quad\quad\quad \forall m, t, \omega.
	\label{eq:db_tdep}
\end{align}
This is a detailed balance condition on the ``reservoir" spectrum seen by the principal spins that ideally should hold at all times. The sufficiency of this condition is clear from \cref{eq:tdep_gibbs}; if it holds all terms in the sum on the right-hand-side are zero. From the Lorentzian form of our engineered time-dependent spectral densities, \cref{eq:lambda}, it is easy to confirm that this detailed balance condition is \emph{not} always fulfilled. In fact, we will see that the thermalization performance of our protocol is intimately linked to the degree to which detailed balance is violated by the engineered spectral densities. 

\begin{figure}[t]
	\includegraphics[scale=0.23] {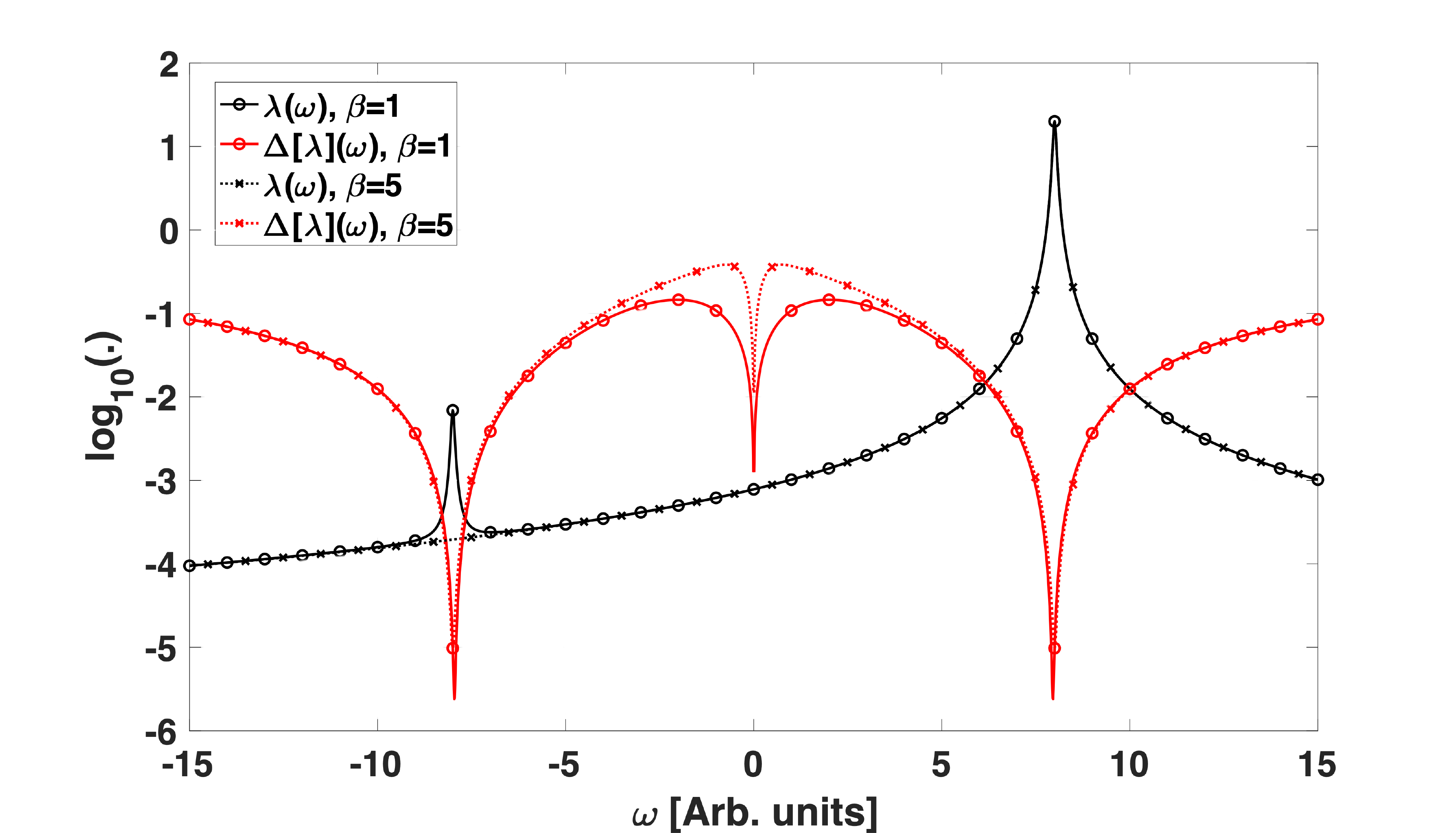}
	\caption{Examples of the engineered spectral density ($\lambda(\omega)$) that dictates transition rates in the dynamical master equation for the principal spins, and the degree to which these rates violate detailed balance ($\Delta[\lambda](\omega)$), as a function of $\omega$. Note that the vertical axis is in log scale. The black curves are $\lambda(\omega)$ and the red curves are $\Delta[\lambda](\omega)$, while the markers and linestyle distinguish between intermediate and low temperature ($\beta=1$ and $\beta=5$, respectively). $\Omega_m(t)=8$ and $\Gamma=0.1$ for all curves.}
	\label{fig:rates_and_db}
\end{figure}

However, before discussing detailed balance further it helps to make another observation about the reduced description of dynamics of the principal spins. The Lindblad form of the master equation in \cref{eq:L} implies that the populations and coherences (on-diagonal and off-diagonal elements of the density matrix, respectively) in the eigenbasis of $H_{\rm sys}$ undergo decoupled dynamics. In fact, assuming for simplicity that $H_{\rm sys}$ has no degenerate energies or energy gaps, and denoting the populations by $P_\epsilon(t) \equiv \bra{\epsilon} \rho(t) \ket{\epsilon}$ and coherences by $c_{\epsilon, \epsilon'}(t) = \bra{\epsilon} \rho(t) \ket{\epsilon'}$, the dynamics of these quantities follows:
\begin{align}
	\dot{P}_\epsilon(t) &= \sum_{m=1}^M g_m^2 \sum_{\epsilon'} \lambda^m_t(\omega_{\epsilon,\epsilon'}) P_{\epsilon'}(t) -  \lambda^m_t(\omega_{\epsilon', \epsilon}) P_{\epsilon}(t)\nonumber \\
	\dot{c}_{\epsilon, \epsilon'}(t) &= -\frac{1}{2}\left( \sum_{m=1}^M g_m^2 \sum_{f} \lambda^m_t(\omega_{f,\epsilon})+\lambda^m_t(\omega_{f,\epsilon'}) \right) c_{\epsilon, \epsilon'}(t). 
	\label{eq:pop_coh_eqs}
\end{align}
Here, $\omega_{\epsilon,\epsilon'}\equiv \epsilon'-\epsilon$, and the sums over $\epsilon'$ in the first line (and $f$ in the second line) are restricted to being only over the allowed transitions; \ie where $X_m(\omega_{\epsilon, \epsilon'})$ is non-zero ($X_m(\omega_{f, \epsilon})$ is non-zero for the second line).
We observe that the coherences decay exponentially and the populations follow a (time-dependent) rate equation. 

Consider a two-state version of the population dynamics with time-independent rates,
\begin{align}
	\dot{P}_0(t) &= \lambda(\omega)P_1(t) - \lambda(-\omega)P_0(t) \nonumber \\
	\dot{P}_1(t) &= \lambda(-\omega)P_0(t) - \lambda(\omega)P_1(t), \nonumber
\end{align}
with $\omega = \epsilon_1-\epsilon_0 >0$.
The steady-state of this dynamics is given by $P_0^{\rm eq}=\frac{\lambda(\omega)}{\lambda(\omega)+\lambda(-\omega)}$ and $P_1^{\rm eq}=1-P_0^{\rm eq}$. This steady-state yields Boltzmann distributed populations if detailed balance is satisfied; \ie $\lambda(-\omega)=\lambda(\omega)e^{-\beta\omega}$. Using this property, we define a metric for violation of detailed balance by transition rates $\lambda(\pm \omega)$, as 
\begin{align}
	\Delta[\lambda](\omega) \equiv \left\vert \frac{\lambda(-\omega)-\lambda(\omega)e^{-\beta\omega}}{(\lambda(\omega)+\lambda(-\omega))(1+e^{-\beta \omega})}\right\vert
	\label{eq:Delta}
\end{align}
This is the total variation distance (TVD) between the ideal Boltzmann distributed populations (at inverse temperature $\beta$) of a two-state system with energy gap $\omega$ and the equilibrium distribution attained by the rates $\lambda(\pm \omega)$. 

This metric helps us understand the detailed balance properties of the engineered transition rates for our protocol given in \cref{eq:lambda}. In \cref{fig:rates_and_db} we plot $\log_{10}(\lambda^m_t(\omega))$ and $\log_{10}(\Delta[\lambda^m_t](\omega))$ as a function of $\omega$ for a fixed time $t$ for which $\Omega_m(t)=8$. Two key points to note from this figure are: (i) as noted previously, the transition rates are significant only in a narrow range around $\pm \Omega_m(t)$, and (ii) violation of detailed balance is minimal at resonance ($\omega=\Omega_m(t)$) and at $\omega\sim 0$, and is significant only away from these regions. This is encouraging because it means that most of the eigenstate population redistribution will occur for transitions around resonance with $\Omega_m(t)$, and in this region the detailed balance violation is small. Therefore, while the lack of detailed balance of our engineered rates will ultimately limit the thermalization performance, we expect that due to the properties highlighted above, the impact will be minimal in many cases. 

It also helps to examine an explicit expression for $\Delta[\lambda](\omega)$, and some of its limiting properties:
\begin{widetext}
\begin{align}
\Delta[\lambda](\omega) &= \frac{2 e^{\frac{1}{2} \beta  (\omega +\Omega )} \left(\sinh \left(\frac{\beta  \omega
   }{2}\right) \cosh \left(\frac{\beta  \Omega }{2}\right) \left(\Gamma^2+4 \left(\omega
   ^2+\Omega ^2\right)\right)-8 \omega  \Omega  \cosh \left(\frac{\beta  \omega
   }{2}\right) \sinh \left(\frac{\beta  \Omega }{2}\right)\right)}{\left(e^{\beta  \omega
   }+1\right) \left(e^{\beta  \Omega }+1\right) \left(\Gamma ^2+4 \left(\omega ^2+\Omega
   ^2\right)\right)} \nonumber \\
	\Delta[\lambda](\omega) &\rightarrow 0  \quad \textrm{as} \quad \omega \rightarrow 0 \nonumber \\
	\Delta[\lambda](\omega) &\rightarrow 0  \quad \textrm{as} \quad \beta \rightarrow 0 \nonumber \\	
	\Delta[\lambda](\omega) &\rightarrow \frac{1}{2} - \frac{4 \omega \Omega_m(t)}{\Gamma^2 + 4(\omega^2 + \Omega_m(t)^2)}  \quad \textrm{as} \quad \beta \rightarrow \infty \nonumber 
\end{align}
\end{widetext}
From the limiting behavior we see that the detailed balance violation reduces for increasing temperature and for decreasing frequencies. Moreover, for decreasing temperature, the error metric limits to a saturating non-zero value, meaning that at low temperatures detailed balance violation will never be negligible, except possibly at resonance ($\omega=\Omega_m(t)$). 

Another interesting observation is the dependence of $\Delta[\lambda]$ on $\Gamma$. For fixed values of the other parameters, we see that decreasing $\Gamma$ decreases $\Delta[\lambda]$ polynomially. However, due to the operating regime outlined in \cref{eq:regime}, a decrease in $\Gamma$ should be accompanied by a decrease in the ancilla sweeping rate ($\vert df_m/dt\vert$) and system-ancilla coupling strength ($g$), meaning that the overall running time of the protocol increases. 

It is worth elucidating the physical reason why $\Gamma \rightarrow 0$ yields higher quality thermalization; the ideal scenario from the system's perspective is to ``see" ancilla degrees of freedom at each instant in time with Boltzmann distributed populations and frequencies that are sharply resonant with transitions in $H_{\rm sys}$. However, the entropy reduction mechanism that prepares the correct population distribution in the ancilla DOF, \eg optical pumping, necessarily broadens the transition energies of the ancilla. Hence, achieving the ideal scenario of sharp resonances simultaneously with Boltzmann distributed populations is impossible with our protocol, and the best one can do is approximate this with $\Gamma \rightarrow 0$.

Based on these observations about the behavior of the engineered spectral density that enters the dynamical master equation for the principal spins, we can surmise that the most challenging scenario for our protocol will be to thermalize to low temperature, $\beta \gg 1$, a system with energy gaps that are close to each other, but not degenerate and not close to zero. In this case the detailed balance violation at frequencies away from resonance -- at the tails of the Lorentzian lineshape of the ancilla -- is large, and due to the closely spaced gaps, there will be system transitions that lie on these tails. Therefore, while eigenstate population redistribution for transitions on resonance with the ancilla DOF (at any time) will mostly satisfy detailed balance, the non-zero spectral density away from resonance will drive nearby (in frequency) transitions, and the resulting population redistribution will violate detailed balance significantly. These conclusions will be illustrated in the numerical simulations in the next section and Appendix \ref{app:parameters}.

\section{Potential Experimental Realization}
\label{sec:sims}
In this section we evaluate the conclusions of the previous section through numerical simulation of the thermalization protocol on a small system that could also form the basis of a minimal experimental realization of the protocol.

We choose a system of two principal spins governed by the Hamiltonian 
\begin{equation}
    H_{\rm sys} = -0.7\sigma_z^1 -B\sigma_z^2 +\left(\sigma_z^1\sigma_z^2+A\left(\sigma_x^1\sigma_x^2+\sigma_y^1\sigma_y^2 \right)\right),
    \label{eq:eg_ham}
\end{equation}
with variable parameters $A$ and $B$.
Each principal spin is coupled to an ancilla with the interaction 
\begin{equation}
    H_{\rm sys-ancilla} = \sum_{m=1}^2 g\left(\sigma_x^m\tau_x^m\right).
\end{equation}
Note that we choose the system-ancilla coupling strength, $g$, to be the same for both ancillae, and also choose the ancilla damping parameter $\Gamma$ and energy sweeping rates to be independent of $m$ (hence we drop the $m$ subscript on various quantities in the following). We choose a $\sigma_x \tau_x$ coupling to illustrate our scheme, but a $\sigma_y\tau_y$ or $\left(\sigma_x\tau_x + \sigma_y\tau_y\right)$ coupling scheme would be equally valid. 

In the simulations below we use a piecewise linear (or sawtooth) sweep of the ancilla energies; \ie over one cycle the modulation looks like $\Omega(t) = (t/T_{\rm cycle})\omega_{\rm max}$ with $0\leq t \leq T_{\rm cycle}$, and $\omega_{\rm max}$ being the largest transition frequency for the system spectrum. The frequency is chosen so as to satisfy the parameter regime in \cref{eq:regime}; \ie $1/T_{\rm cycle} \ll g$.  

In the following, we find it useful to vectorize the dynamical equation in  \cref{eq:L}, and write 
\begin{align}
	\frac{d \vec{\rho}(t)}{dt} = \hat{M}(t)\vec{\rho}(t),	
\end{align}
with
\begin{widetext}
\begin{align}
    \hat{M}(t) = \sum_{m=1}^M g_m^2 
       \sum_\omega \lambda^m_t(\omega)\left[\left(X_m^\dagger(\omega)^T\otimes X_m(\omega) \right) -\frac{1}{2} \left( \mathcal{I} \otimes X_m(\omega)^\dagger X(\omega)\right) - \frac{1}{2}\left(\left(X_m(\omega)^\dagger X_m(\omega)\right)^T\otimes \mathcal{I}\right) \right],
       \label{eq:M}
\end{align}
\end{widetext}
where $\vec{\rho}(t)$ is a vector whose elements are formed by stacking the columns of the density matrix, $\rho(t)$, and $\mathcal{I}$ is the identity matrix of the same dimension as $\rho(t)$.  We can formally solve this explicitly linear equation and write
\begin{align}
	\vec{\rho}(t) \equiv V(t,0)\vec{\rho}(0) = T_\leftarrow[ e^{\int_0^t dt'\hat{M}(t')}]\vec{\rho}(0),
\end{align}
where $T_\leftarrow$ denotes time ordering.

We evaluate the steady-state of the dynamical system determined by our thermalization protocol by computing the first-order Trotter product approximation to the integrated map over a period
\begin{align}
    V(T_{\rm cycle},0) \approx \prod_{i=1}^{T_{\rm cycle}/\Delta t} e^{\Delta t\hat{M}(i\Delta t)},
    \label{eq:map}
\end{align}
with $\Delta t=0.01$. Since the dynamical generator in \cref{eq:M} is periodic, the eigenvalue-zero eigenvector (the kernel solution) of this map over one period is the steady state of the engineered evolution. We denote this steady state as $\rho_{\rm ss}$.

In order to assess the quality of thermalization we compute the trace distance between the density matrix at any time and the ideal thermal state, \ie $\Vert \rho_t - \rho_\beta \Vert$, with $\Vert \mathcal{O} \Vert \equiv \frac{1}{2}\text{tr}\left(\sqrt{\mathcal{O}^\dagger \mathcal{O}}\right) $. The quality of thermalization of the steady state is then $\Vert \rho_{\rm ss} - \rho_\beta \Vert$.

\begin{figure}[t]
	\includegraphics[width = 3in]{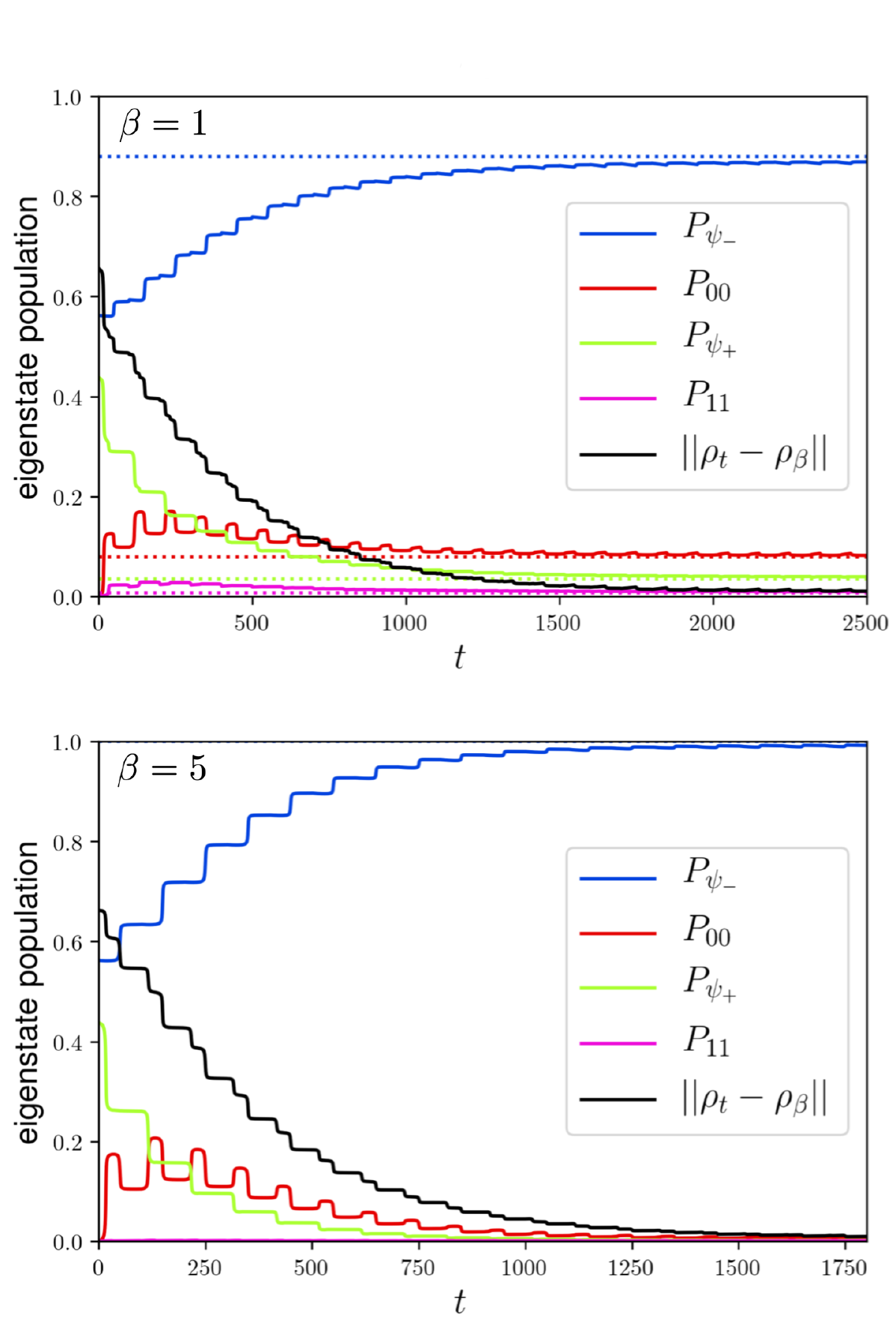}
	\caption{Time evolution of eigenstate populations (blue, red, green, and magenta solid curves) and trace distance to ideal thermal state (black solid curve) for inverse temperature (a) $\beta = 1$ and (b) $\beta=5$. The Hamiltonian parameters are $A=0.8, B=0.5$. The dotted lines show the populations for each of the eigenstates in the ideal thermal state. Other parameters used in the simulations are specified in the main text. 
	\label{fig:pops_t}}
\end{figure}

First, consider a fixed system Hamiltonian, given by $A=0.8$ and $B=0.5$. The eigenstates of this Hamiltonian, in order of increasing energy are $\ket{\psi_-}=\alpha\ket{01}-\beta\ket{10}$, $\ket{00}$, $\ket{\psi_+}=\alpha\ket{01}+\beta\ket{10}$, $\ket{11}$, for real coefficients $\alpha \approx \beta \approx \frac{1}{\sqrt{2}}$. 
To illustrate the thermalization dynamics we propagate the initial state $\ket{01}$ according to the thermalizing master equation \cref{eq:L}, with $\Gamma=0.1, g=0.1$, and the piecewise linear ancilla energy sweeps as mentioned above. 
\cref{fig:pops_t} shows the time development of eigenstate populations and the trace distance to the ideal thermal state under this evolution over a number of consecutive linear sweeps of ancilla energies over the range $[0,\omega_{\rm max}]$. We choose $\omega_{\rm max}\sim 5.7$, which is a little over the maximum energy gap in the system. We show time evolution of the populations for two inverse temperatures, $\beta=1$ and $\beta=5$. All populations monotonically converge to their ideal values for the given Hamiltonian, with $||\rho_t-\rho_\beta|| < .01$ at the end of the shown time evolution. Note that the changes in populations mostly occur in jumps over short time periods, and this is because most of the population mixing dynamics occurs during short periods when an ancilla DOF is in resonance with a system energy gap. 

In Appendix~\ref{app:parameters} we present a thorough analysis of the thermalization performance in terms of the ancilla parameters, desired temperature, and the system Hamiltonian parameters. Although the construction of the thermalization protocol was mostly independent of the system Hamiltonian (except for the range of frequencies over which the ancilla are swept, which is set by the spectral range of $H_{\rm sys}$), we show that properties of the Hamiltonian subtly influence thermalization performance. In particular, the impact of the violation of detailed balance at intermediate and low temperatures can vary according to system Hamiltonian properties. To summarize the findings in Appendix \ref{app:parameters}, the two cases where we find low thermalization performance are: \emph{(i)} if the energy gaps in $H_{\rm sys}$ are congested in frequency (but not degenerate or close to zero), leading to off-resonant transitions driven by rates that significantly violate detailed balance, or \emph{(ii)} at very low temperatures, if some ancilla-induced transitions are thermally suppressed, effectively hindering ergodicity (even though the system is formally ergodic). 

Finally, in Appendix \ref{app:larger_n} we simulate the protocol and evaluate thermalization performance for larger systems, up to four principal spins.

\subsection{Realistic parameters}
\label{sec:parameters}

In order to make the preceding numerical study more concrete we now map the parameters that determine the thermalizaton protocol to realistic values on several quantum simulation platforms. 

In most quantum simulation and quantum computing platforms the most stringent restriction is the strength of achievable spin-spin couplings. We refer to this maximum achievable coupling strength as $J_{\rm max}$ and choose $g = \Gamma = J_{\rm max}/10$ since the system-ancilla coupling must be weaker than the couplings between principal spins. For example, in trapped ions coupling is typically achieved though the M{\o}lmer-S{\o}renson interaction \cite{sorensen_quantum_1999, sorensen_entanglement_2000, milburn_ion_2000}, whose strength depends on several factors including as trap geometry, but typical values are a few kHz (\eg Refs. \cite{kim_quantum_2010, gorman_engineering_2018}) and therefore for this platform we set $J_{\rm max}=10$kHz. Hence, $g=\Gamma=1$kHz. So for example, in the simulations depicted in \cref{fig:pops_t} where we took $g=0.1$, the units of time is $100\mu$s. Finally, a value of $\beta$ determines real thermalization target temperature on this platform through $T=\frac{h(10\textrm{kHz})}{k_B\beta}$, and hence $\beta=1 (\beta=5)$ corresponds to $T\approx 0.48\mu$K ($T\approx 96n$K). Therefore, referring to \cref{fig:pops_t}, we see that the thermalization protocol executed on the trapped ion platform enables generation of a thermal state of the two-spin system governed by Hamiltonian \cref{eq:eg_ham} for values $A=0.8, B=0.5$ at temperature $T\approx 0.48\mu$K ($T\approx 96n$K) in time $T_{\rm th}\approx 0.25$s ($T_{\rm th} \approx 0.18$s). 

We can similarly estimate the real temperatures corresponding to the simulations in \cref{fig:pops_t} and corresponding protocol running times for the superconducting qubit platform and the trapped neutral atom platform, and the results are presented in \cref{tab:table1}. For the superconducting platform we use typical values of cross-resonance gate based couplings of $\sim 4$MHz \cite{Sheldon2016,Chow2011} to fix $J_{\rm max}=10$MHz. For neutral atoms we assume spin-spin coupling through Rydberg interactions, and assuming a Rydberg blockade radius of 10 $\mu$m, where the ratio of the Rydberg interaction to the ground-state interaction is large, we can fix the maximum coupling strength at $J_{\rm max}=2$MHz \cite{RydbergRev_2010,Labuhn_2016}. In addition to the ability to engineer the system Hamiltonian and the system-ancilla couplings, our scheme requires an entropy reduction mechanism like optical pumping. Optical pumping or some variant of it are fairly routine on the neutral atom and trapped ion platforms, and similar mechanisms are also possible in superconducting circuits \cite{you_atomic_2011}.

In all cases presented in \cref{tab:table1}, the protocol runtimes are at the upper limit of typical quantum information protocol runtimes for the given experimental platforms. Hence decoherence mechanisms will impact dynamics, and may even be the dominant source of thermalization errors. Further analysis with particular decoherence models is necessary to understand the interplay between decoherence and the intrinsic open-system evolution engineered by our protocol, and how the former impact thermalization quality. 

Finally, we note that the values presented in \cref{tab:table1} are only for one set of Hamiltonian parameters and for a particular choice for $g$, and mainly serve to demonstrate that the thermalization protocol we have designed is realistic with existing technology. Given more precise estimates of experimentally feasible parameters and a desired $H_{\rm sys}$, it is possible to do a more specific analysis of the achievable thermalization temperatures and runtimes.

\begin{table*}
\caption{\label{tab:table1} Thermalization temperatures ($T$) and protocol runtimes ($T_{\rm th}$) for realistic parameters from three experimental platforms.  }
\begin{ruledtabular}
\begin{tabular}{llllll}
Platform & $J_{\rm max}$ (MHz) & $T$ for $\beta=1$ & $T_{\rm th}$ for $\beta=1$ & $T$ for $\beta=5$ & $T_{\rm th}$ for $\beta=5$ \\
\hline
Trapped Ions & $.01$ & $0.5\mu$K & $0.3$s & $96n$K & $0.2$s \\
Superconducting Qubits & $4$ & $0.2$mK & $0.6$ms & $38\mu$K & $0.44$ms\\
Neutral Atoms & $2$ \footnote{Value extrapolated from exponential curve fit of experimental data \cite{Labuhn_2016}.} & $96\mu$K & $1.3$ms & $19\mu$K & $0.8$ms\\
\end{tabular}
\end{ruledtabular}
\end{table*}

\section{Conclusions}
\label{sec:concl}
Naive approaches to engineering thermalization in quantum simulators require either a coupling to a macroscopic harmonic reservoir or coupling to an extensive number of auxiliary states, each resonant with frequency transitions of the system eigenstates. Both of these approaches require a large number of ancillary DOF, and potentially also a detailed knowledge of the system spectra. In this work, we have developed an alternative approach to engineered thermalization with reduced resource counts by introducing a periodically driven and dissipated ancilla DOF, such that a single ancilla is resonant with different transitions at different times. This time-dependent approach merely requires an estimation of the full spectral width of the system Hamiltonian, and sufficient ancilla DOF to ensure ergodic system dynamics. Considering the interactions between the system and engineered ancilla DOF to be weak compared to interactions within the system, we derived a reduced description of the dissipative dynamics of the system that is in the form of a (time-dependent) Lindblad master equation. We then proved that the thermal state is the fixed point of the system evolution when a detailed balance condition is satisfied by the spectral density generated by the ancilla DOF, and also showed how violations of this condition impact thermalization performance. 

Numerical investigations of a simple lattice system reveal the importance of parameter choices for generating thermal states using this protocol. In particular, the impact of detailed balance violation by our engineered ``reservoir'' varies according to the thermalization temperature and spectral properties of the system Hamiltonian. Thermalizing a many-body system using our protocol is most challenging when the system Hamiltonian has many gaps (transition energies) that are closely spaced (but not exactly degenerate or close to zero) and one demands thermalization to low temperatures. In this regime, the violation of detailed balance becomes most detrimental to thermalization performance. 
It is worth contrasting this with the thermalization of stabilizer Hamiltonians, which does not suffer from detailed balanced violation or increased thermalization time at low temperature because the system gaps (and thus ancilla energies) are constant and known \cite{young_finite_2012,herdman_stroboscopic_2010}.
Numerical investigations also revealed an interesting obstruction to thermalization at low temperatures; a type of degradation of ergodicity due to suppression of certain transitions from lack of thermal energy.

We have developed this thermalization protocol in the framework of continuous-time dynamics. In future work, we will investigate what a discrete-time, or gate-based, version of such a time-dependent ancilla-driven thermalization protocol looks like (\emph{cf.} the spectral combing protocol in Ref. \cite{kaplan_ground_2017}). In particular, we are interested in whether a gate-based version of such a thermalization protocol enables one to overcome the fundamental limitation we identified in the continuous-time protocol; namely, that the entropy reduction mechanism that prepares the correct population distribution in the ancilla DOF also broadens the transition energies of the ancilla, and results in driving of off-resonant transitions at rates that violate of detailed balance. Besides, the formulation of a gate-based version of our protocol would enable rigorous comparison to well-established gate-based thermalization protocols, \eg Ref. \cite{temme_quantum_2011}.

\acknowledgements{Sandia National Laboratories is a multimission laboratory managed and operated by National Technology \& Engineering Solutions of Sandia, LLC, a wholly owned subsidiary of Honeywell International Inc., for the U.S. Department of Energy's National Nuclear Security Administration under contract {DE-NA0003525}. This paper describes objective technical results and analysis. Any subjective views or opinions that might be expressed in the paper do not necessarily represent the views of the U.S. Department of Energy or the United States Government. This material is based upon work supported by the U.S. Department of Energy, Office of Science, Office of Advanced Scientific Computing Research, under the Quantum Computing Application Teams and Quantum Algorithm Teams programs. }

\bibliography{thermostat.bib}

\clearpage
\widetext
\appendix
\section{Derivation of reduced master equation for principal spin lattice}
\label{app:derivation}

As described in the main text, physical motivations led to a choice of parameter regime for the thermalization protocol:
\begin{align}
	\vert\frac{df(t)}{dt}\vert \ll g_m \lessapprox \Gamma^m  \ll ||H_{\rm sys}||, ~~~ \forall m,t
\end{align}

The dynamics of the system (principal spins defining the many-body model and the ancilla spins) are governed by the Hamiltonian in \cref{eq:full_ham} and the dissipative evolution of ancilla spins given in \cref{eq:optical_pumping_me}. Define $R(t)$ as the density matrix for the combined system and ancilla spins -- \ie $R(t) \in \mathcal{B}^+(\mathcal{H}_{\rm sys }\otimes \mathcal{H}_{\rm ancilla})$, the space of positive, trace one operators on the combined Hilbert space. Further, let $\rho(t)=\tr_{\rm a}\{R(t)\} \in \mathcal{B}^+(\mathcal{H}_{\rm sys })$ be the reduced density operator for the system ($\tr_{\rm a}$ denotes a trace over the ancilla subsystem).

We will derive a reduced master equation for $\rho(t)$ assuming one ancilla spin ($M=1$). The result generalizes immediately since each ancilla is weakly coupled and is independent of all other ancilla, and hence additivity of Lindblad evolutions applies \cite{kolodynski_adding_2018}. In the following, we will follow the standard derivation of a master equation in the Born and Markov approximations, but we need to justify some of the steps carefully in this context since instead of a macroscopic reservoir we have a damped and driven ancilla spin.

In the interaction picture (with respect to the free evolution of the system and ancilla) the evolution of the combined system is
\begin{align}
	\frac{d\tilde{R}(t)}{dt} = -i[\tilde{H}_I(t),\tilde{R}(t)] = -ig[\tilde{\sigma}^k_x \tilde{\tau}_x, \tilde{R}(t)],
\end{align}
for some spin in the system, $k$, and where the tilde denotes operators in the interaction picture.
Following the standard derivation \cite{breuer_theory_2002}, by iterating this equation and formally integrating we get
\begin{align}
	\frac{d\tilde{\rho}(t)}{dt} = -\int_0^t ds \tr_{\rm a} \{[\tilde{H}_I(t), [\tilde{H}_I(t-s),\tilde{R}(t-s)]]\}.
\end{align}
At this point we make three approximations. First, we replace $\tilde{R}(t-s)$ with $\tilde{\rho}(t-s)\otimes r^{\rm eq}_{t-s}$, with the justification being that in the weak coupling limit, any entanglement between system and ancilla is quickly damped by the ancilla dynamics and hence we can replace the joint state with a tensor product state. Moreover, the state of the ancilla will can be approximated by an equilibrium state $r^{\rm eq}_{t-s}$, which we will take later to be the thermal state with respect to some $\beta$ and the transition energy at time $t-s$, $\Omega(t-s)$. Note that in standard derivations of the Born-Markov master equation, this state is taken to be time-independent, whereas since we need to sweep the ancilla energy, we keep the time-dependence on the ``reservoir'' state. The second approximation is that we will replace $\tilde{\rho}(t-s)\otimes r^{\rm eq}_s$ with $\tilde{\rho}(t)\otimes r^{\rm eq}_{t}$, with the justification that the trace over the ancilla degrees of freedom will yield a correlation function that decays rapidly as $s\rightarrow t$, and hence only the values of the integral around $s\approx 0$ really contribute. Finally, we take the upper limit of the integral $t\rightarrow \infty$, again using the approximation that the correlation function decays rapidly and so integrand values at large values of $s$ are negligible. With these approximations we get:
\begin{align}
	\frac{d\tilde{\rho}(t)}{dt} = -\int_0^\infty ds \tr_{\rm a} \{[\tilde{H}_I(t), [\tilde{H}_I(t-s),\tilde{\rho}(t)\otimes r^{\rm eq}_{t}]]\}.
		\label{eq:formal_me}
\end{align}

Now let us examine the form of $\tilde{H}_I$ more closely. 
We first define a frequency resolved operator on the system
\begin{align}
	X(\omega) = \sum_{\{\epsilon',\epsilon | \epsilon'-\epsilon = \omega\}}\Pi(\epsilon) \sigma_x^k \Pi(\epsilon'),
\end{align}
where $\Pi(\epsilon)$ is a projector on the eigenspace of $H_{\rm sys}$ with eigenvalue $\epsilon$. Using this definition, the following properties are easy to prove:
\begin{align}
e^{iH_{\rm sys}t} X(\omega) e^{-iH_{\rm sys}t} &= e^{-i\omega t} X(\omega) \nonumber \\
e^{iH_{\rm sys}t} X\dg(\omega) e^{-iH_{\rm sys}t} &= e^{i\omega t} X\dg(\omega) \nonumber \\
X\dg(\omega) &= X(-\omega) \nonumber \\
\sum_\omega X(\omega) &= \sum_\omega X\dg(\omega) = \sigma_x^k \nonumber \\
[H_{\rm sys}, X\dg(\omega)X(\omega)]&=0.
\label{eq:X_props}
\end{align}

Using these properties, we expand \cref{eq:formal_me} as:
\begin{align}
\frac{d\tilde{\rho}(t)}{dt} = g^2\sum_{\omega, \omega'} e^{i(\omega'-\omega)t} \Lambda_t(\omega) 
\Big( X(\omega) \tilde{\rho}(t) X\dg(\omega') - X\dg(\omega')X(\omega) \tilde{\rho}(t) \Big) + h.c.,
\end{align}
with 
\begin{align}
\Lambda_t(\omega) = \int_0^\infty ds e^{i\omega s} \tr\{ \tilde{\tau}_x(t) \tilde{\tau}_x(t-s) r^{\rm eq}_t \}.
\label{eq:corr_fxn}
\end{align}
Assume that over sufficiently long times (since we are mostly concerned about steady state properties long times are of primary interest) the oscillating factor damps any terms with $\omega \neq \omega'$. Dropping these terms (which amounts to the secular, or rotating wave, approximation), we get:
\begin{align}
\frac{d\tilde{\rho}(t)}{dt} = g^2\sum_{\omega} \Lambda_t(\omega) 
\Big( X(\omega) \tilde{\rho}(t) X\dg(\omega) - X\dg(\omega)X(\omega) \tilde{\rho}(t) \Big) + h.c.
\end{align}

Now let us examine the Fourier transforms of the reservoir correlation function, \cref{eq:corr_fxn}. As is customary in the derivation of a Lindblad master equation we assume that the correlation functions are homogeneous in time. This follows easily for large reservoirs near equilibrium but needs more careful thought in our case where we have a driven, dissipated ancilla spin. Since the ancilla energies are periodic there is clearly some non-homogeneity to the correlation function. However, since the change in energy is much slower than the system-reservoir coupling and the ancilla damping we assume that on timescales resolved by the interaction the homogeneity of the correlation function is a valid assumption.

Given time-homogeneity of the correlation function we rewrite $\Lambda_t(\omega)$ as
\begin{align}
\Lambda_t(\omega) = \int_0^\infty ds e^{i\omega s} \tr\{ \tilde{\tau}_x(s) \tilde{\tau}_x r^{\rm eq}_t \}.
\label{eq:corr_fxn2}
\end{align}
To evaluate this quantity, we need to know time dynamics of the operator $\tilde{\tau}_x(t)$. These dynamics are dominated by the driving and dissipation of the ancilla spins, and hence we calculate the correlation functions by formulating the adjoint version of the ancilla dynamics from its free Hamiltonian and the master equation in \cref{eq:optical_pumping_me} 
\begin{align}
	\frac{d \varpi (t)}{dt} = i[\varpi(t), \frac{\Omega(t)}{2}\tau_z] + \mathcal{L}\dg_m(t)[\varpi(t)] = \gamma_+(t)\mathcal{D}\dg[\tau_+^m]\varpi(t) + \gamma_-(t)\mathcal{D}\dg[\tau_-^m]\varpi(t) ,
\end{align}
where $\varpi$ is any operator acting on $\mathcal{H}_{\rm ancilla}$, and $\mathcal{D}\dg[A]B \equiv A\dg B A - \frac{1}{2}A\dg A B - \frac{1}{2} B A\dg A$. According to this adjoint evolution equation,
\begin{align}
\dot{\tilde{\tau}}_+(t) &= i\Omega_t \tilde{\tau}_+(t) -\frac{\Gamma_t}{2}\tilde{\tau}_+(t), \nonumber \\\dot{\tilde{\tau}}_-(t) &= i\Omega_t \tilde{\tau}_-(t) -\frac{\Gamma_t}{2}\tilde{\tau}_-(t),
\end{align}
where $\Gamma_t = \gamma_+(t) + \gamma_-(t)$. In the quasi-static limit where we ignore the time-dependence of $\Omega_t$ and $\Gamma_t$, we can solve these equations easily and get
\begin{align}
\tilde{\tau}_x(t) = \tilde{\tau}_+(t) + \tilde{\tau}_-(t) = e^{\left(i\Omega_t - \frac{\Gamma_t}{2}\right)t} \tau_+ + e^{\left(-i\Omega_t - \frac{\Gamma_t}{2}\right)t} \tau_- .
\end{align}
Finally, since $r^{\rm eq}_t$ is a Gibbs state and hence diagonal in the computational basis, we get
\begin{align}
\Lambda_t(\omega) &= \int_0^\infty ds e^{i\omega s} \tr\{ \tilde{\tau}_x(s) \tilde{\tau}_x r^{\rm eq}_t \} \nonumber \\
&= \frac{P^0_t}{\frac{\Gamma_t}{2}-i(\omega - \Omega_t)} + \frac{P^1_t}{\frac{\Gamma_t}{2}-i(\omega + \Omega_t)},
\label{eq:Lambda_final}
\end{align}
where $P^0_t = \bra{0} r^{\rm eq}_t \ket{0}$ and $P^1_t = \bra{1} r^{\rm eq}_t \ket{1}$ are the Boltzmann distributed populations of the ancilla spin.

At this point, we have a Markov master equation (with time-dependent coefficients) describing the evolution of the system alone (in the interaction picture). To further isolate the dynamics of interest, we separate out the Lamb shift resulting from the coupling to the reservoir from the incoherent dynamics. To do this, we separate the real and imaginary components of \cref{eq:Lambda_final}, as $\Lambda_t(\omega) = \frac{1}{2} \lambda_t(\omega) + iS(\omega)$. The Lamb shift Hamiltonian is then \cite{breuer_theory_2002}
\begin{align}
H_{LS} = g^2\sum_\omega S(\omega) X\dg(\omega) X(\omega).
\end{align}
Owing to the properties in \cref{eq:X_props}, we know $[H_{\rm sys}, H_{LS}]=0$, and therefore for the purposes of preparing thermal states of the system, the Lamb shift is inconsequential. It will shift some of the system energies around, however, since our protocol does not require precise knowledge of the system energies this is not important. So we ignore it from now on. The real part of $\Lambda_t$, on the other hand, yields the incoherent dynamics that we want to tune to 
achieve thermalization of the system. Explicitly, the real part is:
\begin{align}
\lambda_t(\omega) = \frac{P^0_t \left(\frac{\Gamma_t}{2}\right)}{\left(\frac{\Gamma_t}{2}\right)^2 + (\omega - \Omega_t)^2} + \frac{P^1_t \left(\frac{\Gamma_t}{2}\right)}{\left(\frac{\Gamma_t}{2}\right)^2 + (\omega + \Omega_t)^2}
\end{align}

To summarize, we write the relevant dynamics of the system alone, in the interaction frame with respect to $H_{\rm sys}$ as:
\begin{align}
\frac{d\tilde{\rho}(t)}{dt} = g^2\sum_{\omega} \lambda_t(\omega) 
\Big( X(\omega) \tilde{\rho}(t) X\dg(\omega) - \frac{1}{2}X\dg(\omega)X(\omega) \tilde{\rho}(t) - \frac{1}{2} \tilde{\rho}(t)X\dg(\omega)X(\omega) \Big)
\end{align}

In the general case where we have many ancilla spins, one simply has a sum of the above Lindblad evolution for each system-ancilla coupling.

\section{Thermal state accuracy as a function of parameters}
\label{app:parameters}

In this Appendix we study the accuracy of thermal state preparation for the two-spin principal system studied in the main text, as a function of the protocol's and system's parameters. 

\begin{figure}
	\includegraphics[scale=0.25]{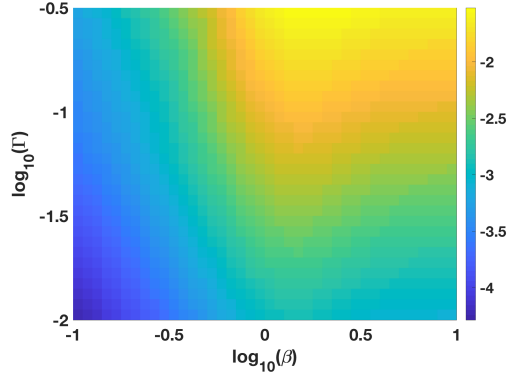}
	\caption{Log of the trace distance between the steady-state of \cref{eq:L} and the ideal thermal state, $\log_{10}\vert\vert \rho_{\rm ss} - \rho_\beta\vert\vert$, for a system Hamiltonian with parameters $A=0.8$, $B=0.5$. See main text for other simulation parameters.
	\label{fig:vary_bath_params}}
\end{figure}

First, we study the behavior as a function of the engineered bath parameters; the ancilla damping rate, $\Gamma$, and the desired inverse temperature, $\beta$. \cref{fig:vary_bath_params} shows the trace distance between the steady-state and the ideal thermal state as a function of these parameters, with the Hamiltonian parameters fixed at $A=0.8, B=0.5$. For each point, we choose the other parameters consistent with the regime in \cref{eq:regime} -- \ie $g=\Gamma$ and $T_{\rm cycle} = 4/\Gamma$. We see that the protocol is able to prepare states that are very close to the true thermal state. The poorest performance is in the region where $\Gamma$ is too large or the desired temperature is too low, but even here the trace distance to the ideal thermal state is close to $10^{-2}$. The region of poorest performance is consistent with our analysis in the main text that showed that the violation of detailed balance increases when $\Gamma$ or $\beta$ increase.

Next, we study thermalization performance as a function of the parameters $A$ and $B$ in $H_{\rm sys}$. Although the construction of the thermalization protocol was mostly independent of the system Hamiltonian (except for the range of frequencies over which the ancilla are swept, which is set by the spectral range of $H_{\rm sys}$), we will see that properties of the Hamiltonian subtly influence thermalization performance. In particular, we show how the impact of the violation of detailed balance at intermediate and low temperatures can vary according to system Hamiltonian properties.

\begin{figure*}
	\includegraphics[width=\textwidth]{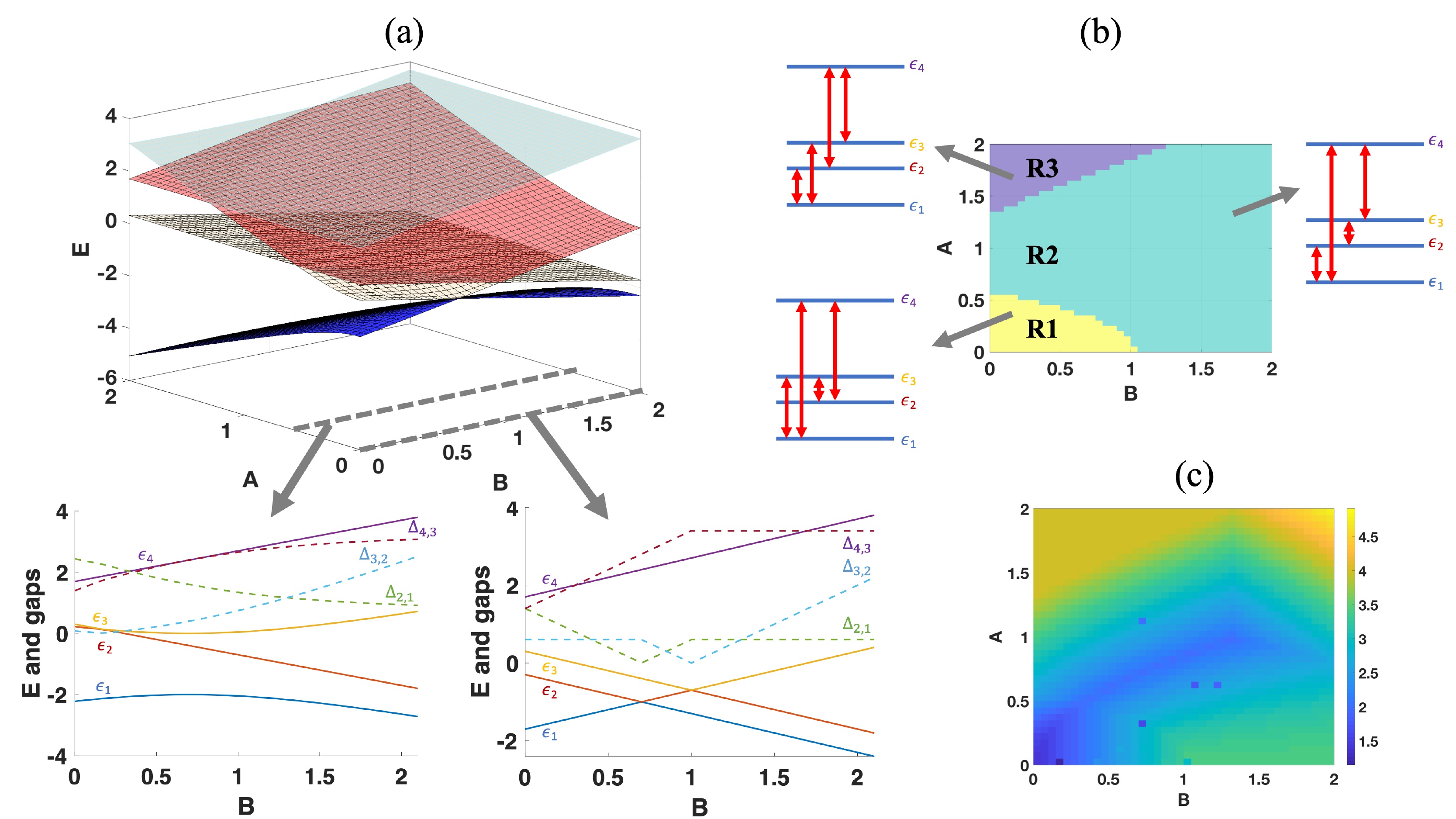}
	\caption{Eigenstate energies and connectivity for two-qubit spin system studied in section \cref{sec:sims}. \textbf{(a)} Energies of the four eigenstates as a function of the Hamiltonian parameters $A$ and $B$. The 2-dimensional plots at the bottom show slices across these energy surfaces at $A=0$ and $A=0.5$. These slices show greater detail and also allow us to plot the energies (solid lines) and gaps between energies (dotted lines). $\epsilon_i$ is the energy of state $i$, and $\Delta_{i,j}=\epsilon_j-\epsilon_i$. \textbf{(b)} The physical coupling to ancilla DOF is through interactions of the form $\sigma^i_x \otimes \tau^i_x$. These result in connectivity (matrix elements) between certain eigenstates of the Hamiltonian $H_{\rm sys}$, which is roughly of three types, depending on the region in $A,B$ parameter space, indicated by the three color regions in this figure, and labeled R1, R2, R3. For each region we indicate the eigenstates that are connected with red arrows. \textbf{(c)} To summarize how the spread in transition energies varies, we show the interquartile range (IQR) for the allowed transitions at each $A,B$ point.}
	\label{fig:energies_and_connectivity}
\end{figure*}

In \cref{fig:energies_and_connectivity} (a) and (b) we show, as a function of the parameters $A$ and $B$, the spectrum of $H_{\rm sys}$ and the connectivity of the eigenstates of $H_{\rm sys}$, respectively. From \cref{fig:energies_and_connectivity} (a) we see that the spectrum varies in a complicated manner across this parameter regime -- importantly, in the region $A \lesssim 0.5$, the spectrum becomes congested, with even several points of degeneracy between states $\epsilon_1$ and $\epsilon_2$, and also $\epsilon_2$ and $\epsilon_3$ (we will label the eigenstates of $H_{\rm sys}$, in order of increasing energy, as $\epsilon_1, \epsilon_2, \epsilon_3, \epsilon_4$, and gaps as $\Delta_{i,j}\equiv \epsilon_j-\epsilon_i$).  The connectivity of the eigenstates can mostly be classified into three types, depending on the region in $A,B$ parameter space. In \cref{fig:energies_and_connectivity} (b) we show these three regions, labeled R1, R2, R3, and how the eigenstates of $H_{\rm sys}$ are connected there -- a red arrow between the states separated by $\omega$, indicates that  $X_m(\omega)$ is non-zero for $m=1$, $m=2$, or both. \cref{fig:energies_and_connectivity} (b) shows that while the system is ergodic for all values of $A,B$, the direct connectivity of eigenstates varies, and this will play an important role in thermalization quality. Finally, in \cref{fig:energies_and_connectivity}(c) we summarize how the spread in transition energies varies for the Hamiltonian parameters by plotting the interquartile range (IQR) for energies of the \emph{allowed} transitions at each $A,B$. We choose IQR to represent the spread of gaps because it is less sensitive to outliers (there is one large gap at most parameter values) than other measures of deviation.

\begin{figure}
	\includegraphics[scale=0.7]{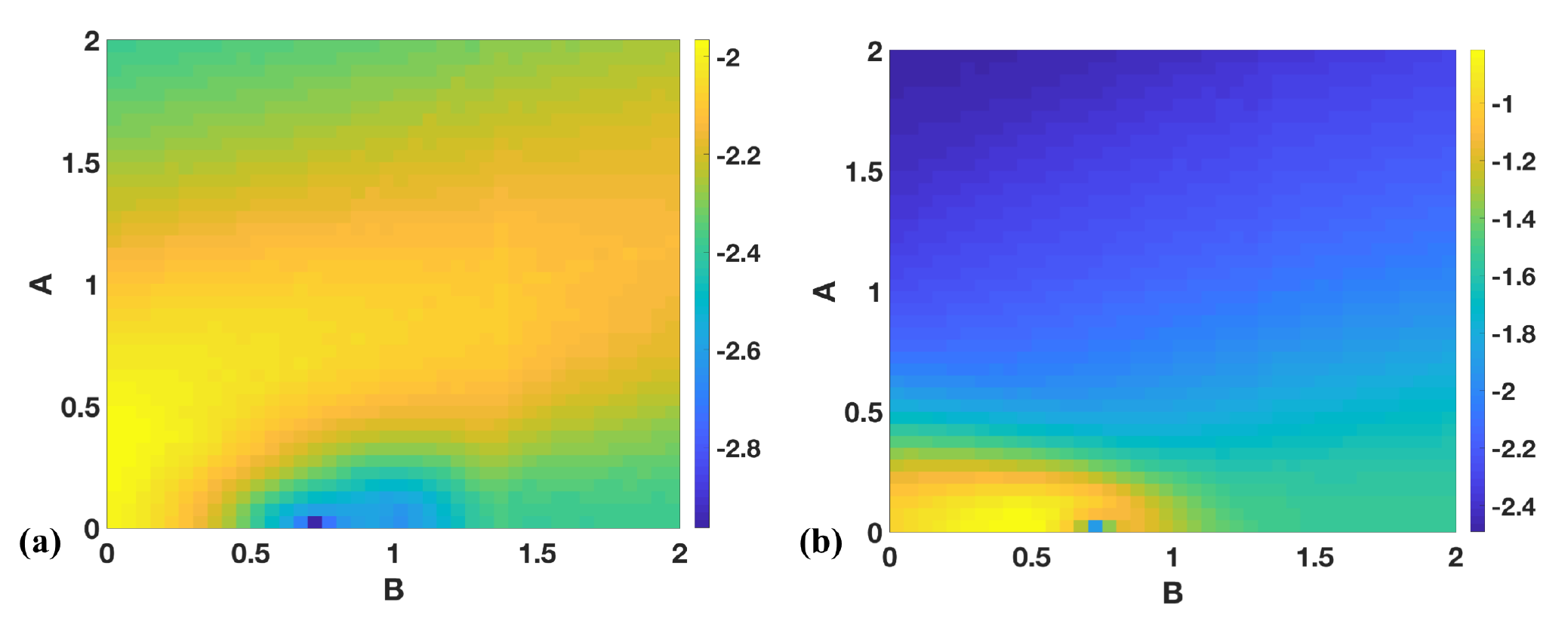}
	\caption{Log of the trace distance between steady state of engineered dynamics and true thermal state, $\log_{10}\vert\vert \rho_{\rm ss} - \rho_\beta\vert\vert$, as a function of Hamiltonian parameters $A$ and $B$ for the model studied in \cref{sec:sims}. \textbf{(a)} Intermediate temperature $\beta = 1$, and \textbf{(b)} low temperature $\beta = 5$.
	\label{fig:vary_ham_params}}
\end{figure}

To summarize thermalization behavior, in \cref{fig:vary_ham_params} we show the trace distance between the ideal thermal state and the steady state of the engineered evolution for intermediate and low temperatures, as a function of the Hamiltonian parameters $A, B$. The values of the other parameters for all of these plots are $\Gamma=g=0.1$, and as before, we used a linear sweep of the ancilla energies over the range $[0,\omega_{\rm max}=1.2\times \Delta_{\rm max}]$, where $\Delta_{\rm max}$ is the maximum gap in the Hamiltonian energies for the chosen $A,B$. The linear sweep duration is set by $T_{\rm cycle}=4/\Gamma$. We restrict to positive values of $A$ because the thermalization performance is symmetric about $A=0$. 

In the intermediate temperature case, \cref{fig:vary_ham_params}(a), we see that the trace distance is below $10^{-2}$ across the whole parameter range. The worst performance is in a wide region in the center of the parameter space, and by comparing \cref{fig:vary_ham_params}(a) to \cref{fig:energies_and_connectivity}(c) we see that this region corresponds to the region with the smallest spread in gaps in the system. Therefore we conclude that the poorer performance in this region is due to the effect identified in \cref{sec:db}. Namely, when there are closely spaced gaps, off-resonant transitions can be driven by rates that significantly violate detailed balance, because as shown in \cref{fig:rates_and_db} around the tails of the instantaneous Lorentzian spectral density, transition rates are non-negligible \emph{and} there is significant detailed balance violation at once. 
Note that in the region $A\sim 0$, $0.5\lesssim B \lesssim 1.2$ the thermalization performance is better even though the gaps are very congested here also. The reason is that the two most important gaps at this temperature, $\Delta_{2,1}$ and $\Delta_{3,2}$, are almost zero and almost degenerate in this region (see \cref{fig:energies_and_connectivity}(a)), and thus the detailed balance violation is softened here (recall that $\Delta[\lambda](\omega)$ decreases around $\omega\sim 0$, and moreover, if two gaps are almost degenerate, then although one may drive the off-resonant transition, it will be close enough to resonance that the violation of detailed balance is minimal).

Moving onto the low temperature case, \cref{fig:vary_ham_params}(b), we see that the trace distances are uniformly greater than the intermediate temperature case. This is consistent with our previous observation that the effectiveness of the thermalization protocol degrades as the target temperature reduces. Despite this, the achieved trace distance at low temperature ($\beta=5$) over most of the Hamiltonian parameter space is fairly small ($\sim 10^{-2}$). The exception to this is in part of region R1, with $A \lesssim 0.3$ and $B \lesssim 1$. To explain the deviation we return to \cref{fig:energies_and_connectivity} (b), where we showed that in region R1 states $e_1$ and $e_2$ are not directly connected by an ancilla-induced transition. At $\beta=5$ the majority of the population of the ideal thermal state in this region is distributed between states $e_1$ and $e_2$, however, if these states are not directly connected, in order to distribute population between these states, one needs to go through transitions to higher energy states. But such transitions are highly suppressed at low temperatures because the thermal energy provided by the ancilla DOFs is too small; \ie $\lambda_t(\omega)\ll 1$ for $\omega \gg 1/\beta$. This accounts for the large trace distance between the steady state and the ideal thermal state in this parameter region of the Hamiltonian at low temperature. This example demonstrates that even if a system is formally ergodic, ideal thermalization can be prevented if all transitions are not thermally activated.

\section{Evaluating thermalization quality for larger system sizes}
\label{app:larger_n}

\begin{figure}[t]
	\includegraphics[width = 4in]{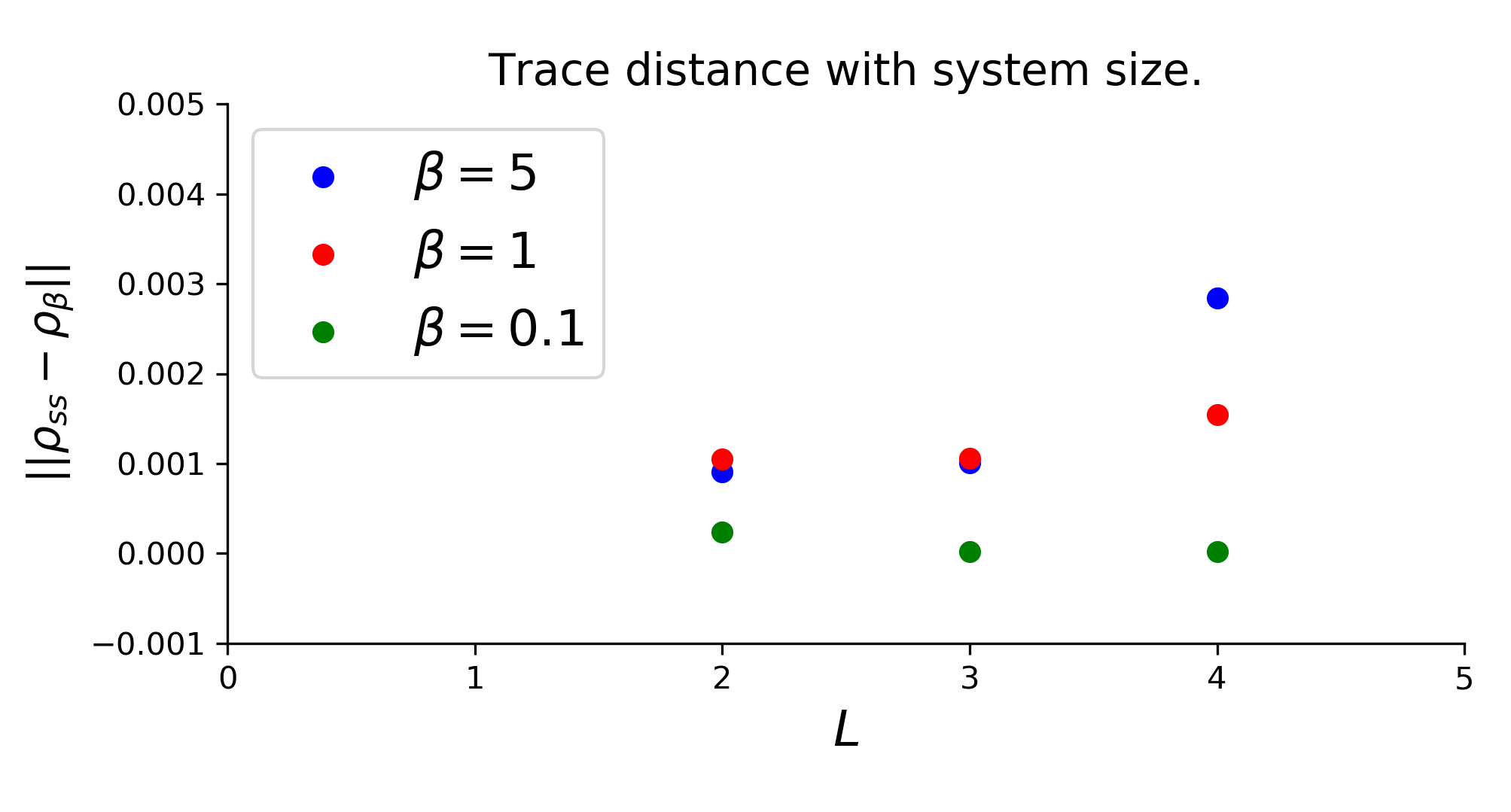}
	\caption{Trace distance between steady-state of engineered dynamics and genuine thermal state of one-dimensional spin chain with length $ 2\leq L \leq 4$ at an effective inverse temperature $\beta = 5, 1, 0.1$. 
	\label{fig:scaling}}
\end{figure}

We evaluate the effectiveness of our protocol numerically by determining the fixed point of evolution for a spin chain with pair-wise interactions with increasing length. The Hamiltonian defines a system with alternating onsite energies and interactions between the X, Y and Z spin degrees of freedom,
\begin{equation}
    H = \sum_{i \in even} B \sigma_z^i + \sum_{i \in odd} \frac{B}{2} \sigma_z^i + \sum_{\langle ij \rangle} J\left[A\left(\sigma_x^i\sigma_x^j + \sigma_y^i\sigma_y^j\right) + \sigma_z^i\sigma_z^j \right].
\end{equation}
Numerically, we build the generator in Equation~\ref{eq:M} and the kernel solution by diagonalizing the time-ordered linear map within Equation~\ref{eq:map} which is equivalent to the steady-state solution $\rho_{ss}$. Coupling alternating principal spins to an ancilla spin, $(M = L/2)$ for even L and $M=(L+1)/2$ for odd L, on the spin chain has a unique steady state solution approximating the thermal state and reduces computational resources required for exact numerical simulation. We compute the trace distance between the steady-state solution from the true thermal state for a one-dimensional spin chain with $B=J=1$ and $A = 0.8$ at varied temperatures Fig.~\ref{fig:scaling}. It is evident the steady-state remains a good approximation of the true thermal state with increasing system size at low, high and intermediate temperatures.

In the numerical algorithm we store the entirety of Equation~\ref{eq:M} in memory to reduce the computational complexity in time, and the amount of memory bytes required to construct Equation~\ref{eq:M} scales $R = \mathcal{O}\left(MN_\omega2^{4L}\right)$ where M is the total number of ancilla spins, $N_\omega$ is the number of unique system frequencies, and L is the length of the spin chain. The computational overhead limits exact numerical simulation to small spin chains, however the generator is sparse and sparse linear algebra computations can reduce memory requirements. Moreover, as shown in Eq. \ref{eq:pop_coh_eqs} in the main text, one could simplify further by just simulating population dynamics. However, in this work we limit ourselves to small-scale simulations designed to demonstrate the basics of the protocol, and do not pursue simulations of large-scale systems.

\end{document}